\documentclass{article}
\usepackage{amsmath,lscape}
\usepackage{natbib}
\usepackage{amssymb}
\usepackage{wasysym}
\usepackage{dcolumn}
\usepackage{pdflscape}
\usepackage{booktabs}
\usepackage{placeins}
\usepackage{xr}

\usepackage{color}
\usepackage[T1]{fontenc}
\usepackage{a4wide}
\usepackage{supertabular}
\usepackage{multirow}
\usepackage[lmargin=2.0cm, rmargin=2.0cm,tmargin=2.5cm,bmargin=2.5cm]{geometry}
\usepackage{setspace} 
\usepackage{graphicx,color} 
\usepackage{graphics}
\usepackage{subfig}
\usepackage{longtable}
\usepackage{ae}
\usepackage{rotating}
\usepackage{url}

\definecolor{Dark}{gray}{.20}

\usepackage{amsmath,amsthm}

\usepackage{hyperref} 
\hypersetup{
  colorlinks = true,
  linkcolor=blue,   
  citecolor=blue,   
  urlcolor=blue,    
  pagebackref=true,
  implicit=false,
  bookmarks=true,
  bookmarksopen=true,
  pdfdisplaydoctitle=true
}

\newcommand{\E}{\mathbb {E}}

\renewcommand{\P}{\mathbb{P}}

\newcommand{\R}{\mathbb {R}}

\newcommand{\var}{\mathrm{Var}}

\newtheorem{theorem}{Theorem}

\begin{document}
\setlength{\baselineskip}{12pt}
\pagestyle{plain}

\begin{center}
\Large \sc U-statistical inference for hierarchical clustering \\[1cm]
\end{center}
\begin{center}\sc Marcio Valk\footnote{Department of Statistics, Federal University of Rio Grande do Sul, Porto Alegre, RS.} \\[0.5cm]

 Gabriela Cybis\footnote{Department of Statistics, Federal University of Rio Grande do Sul, Porto Alegre, RS.} 

\end{center}
\begin{center}
May, 2018\\[.4in]
\end{center}

\noindent\bf Abstract. \rm 
Clustering methods are a valuable tool for the identification of patterns in high dimensional data with applications in many scientific problems. However, quantifying uncertainty in clustering is a challenging problem, particularly when dealing with High Dimension Low Sample Size (HDLSS) data.  We develop here a U-statistics based clustering approach that assesses statistical significance in clustering and is specifically tailored to HDLSS scenarios. These non-parametric methods rely on very few assumptions about the data, and thus can be applied a wide range of dataset for which the euclidean distance captures relevant features. We propose two significance clustering algorithms, a hierarchical method and a non-nested version. In order to do so, we first propose an extension of a relevant U-statistics and develop its asymptotic theory. Our methods are tested through extensive simulations and found to be more powerful than competing alternatives. 
They are further showcased in two applications ranging from genetics to image recognition problems. \\

\noindent \bf JEL Classifications. \rm \\[.1in]
\clearpage

\setlength{\baselineskip}{18pt}
\pagestyle{plain}
\setcounter{page}{2}
\setcounter{footnote}{0}
\clearpage

\section{Introduction}\label{intro}
Identifying systematic patterns in high dimensional data has been a challenge for wide range of subjects, such as econometrics, bioinformatics and genetics.  
Clustering is often a first step in the analysis of multivariate data, dividing the data into sub-populations to reveal meaningful features or populational structure. 
A critical issue is how to determine whether the clusters represent in fact an important feature or are simply the result of the sample variation. In other words, are the clusters statistically significant?  We present here a U-statistics based approach that clusters the data while assessing significance of such partitions.

While most traditional clustering methods require à priori definition of the number of groups $k$ in which the data should be partitioned,  hierarchical clustering procedures are not constrained by such criteria, producing multiple layered partitions. These methods have become increasingly popular in exploratory analysis for many fields since they allow for different structural representations at the different levels. However, it generally falls to heuristic criteria or the researcher's judgement  to define which partition levels should be assigned meaning. By assessing significance in hierarchical clustering, we can identify which clustering layers represent actual populational structure and which are simple consequence of spurious random effects. Thus, in this paper we also develop a hierarchical clustering procedure informed by statistical significance and tailored to high dimentional low sample size data (HDLSS).

Statistical significance in clustering can be assessed by several approaches, for example by considering mixture models of distributions such as the Gaussian \citep{mclachlan2004}. However, this problem becomes increasingly challenging when the data are high dimensional and have small sample sizes, since it requires complete parametric estimation, usually involving costly matrix inversions. 
\cite{mcshane2002} addresses this issue by reducing the dimensionality of a microarray hierarchical clustering problem by considering solely the first few principal components of the data matrix.
\cite{shimodaira2004}  propose an approach inspired on the bootstrap strategy used in phylogenetics to assess confidence in hierarchical clustering, which is implemented in the R package pvclust \citep{suzuki2006}. \cite{maitra2012} assess the significance of a particular cluster arrangement by employing a bootstraping strategy on multiple elipsoidal compact clusters of multivariate data, however the method is not particularly well suited for the HDLSS setting. 
In a more general approach specifically tailored to the HDLSS scenario, \citet{Liu2008} propose a statistical test to assess the significance of clustering the data into $K$ groups, that has been implemented in the R package SigClust.  Their method has been further developed in \citet{Huang2015} to incorporate soft thresholding for critical eigenvalue estimation. Additionally, \citet{Kimes2017} extend the method to assess significance in hierarchical clustering. 
The approach's null hypothesis, however, is that the data come from a single multivariate normal distribution, which can be an issue since rejection of the null may be a simple consequence of non-normal data.

A promising approach to address the same problem is presented in \cite{Cybis2016}, were we develop a U-statistics based framework for clustering and classification that is particularly appealing in the HDLSS scenario prevalent in genetics. Our methods are based on the U-statistic $B_n$ which measures the difference of within and between group dissimilarities in a particular partition of the sample into two groups.
 \citet{Pinheiro2009} show that this statistic belongs to a general class of first order degenerate U-statistics for which asymptotic results can be derived. These asymptotic properties hold, even without assumptions of stochastic independence or homogeneity of the marginal probability laws. Building upon these results, in \cite{Cybis2016} we propose an efficient procedure for testing overall sample homogeneity based on a combinatorial criteria suggested by \citet{Valk2012} in which a sample is considered homogeneous if all partitions into two subgroups are not statistically significant. This approach presents obvious potential for extension to clustering problems, since homogeneity is tested by verifying whether the partition that better separates the samples into two groups is significant. In this paper we explore this U-statistics framework to develop a hierarchical clustering method informed by statistical significance. In order to do so, we must first extend the homogeneity test to construct a standalone significance clustering algorithm. 
 
Critically, our methods do not rely on particular distributional assumptions about the data, being thus applicable to the wide range of problems. However, since the method relies on the Euclidean distance, in should only be applied in contexts in which the data or a transformation thereof is in a space for which the Euclidean distance captures relevant features for group separation.

The steps to developing our significance hierarchical clustering method are outlined as follows. First, in Section 2.1 we present the homogeneity test of \cite{Cybis2016} and its theoretical basis.  The original definition of the $B_n$ statistic requires that both groups have at least two elements, so in order devise a hierarchical clustering algorithm that can properly identify outlier elements, we must extend its definition to contemplate groups of size one. This extensions, together with investigation of theoretical properties that show its compatibility with the previous framework and asymptotic theory are presented in Section 2.2. Subsequently, we note that under this new definition of $B_n$ the homogeneity test does not adequately control type I error for a range of dimensions $L$ and sample sizes $n$. Since this test is based on the maximum of many $B_n$ statistics, in Section 2.3 we explore extreme value theory to increase the range of problems for which our method can be properly applied. Next we consider the issue of clustering based on the newly extended homogeneity test. In section 2.4 we define the Uclust method which finds the statistically significant data partition that better separates the sample into two groups. Finally, in section 2.5 we present our hierarchical clustering algorithm Uhclust, which iteratively applies Uclust to the data while controlling the family-wise error rate (FWER). 

The remainder of the paper focuses on evaluating the methodology through simulation studies, in Section 4, and three applications to real data in Section 5. The applications showcase the method's versatility in two different scenarios.

\section{Methods}

\subsection{U-Statistics based test for group homogeneity}

Let $\boldsymbol{X}=(X_{1}\dots,X_{n})$ be a random sample of $n$ $L$-dimensional vectors and assume that there are  2 groups $G_1$ and $G_2$ of sample sizes $n_1$ and $n_2$, respectively, where $n=n_1+n_2$. In the $g$-th group, observations $X_{g1}\dots,X_{gn_g}$ are assumed to be independent and identically distributed (i.i.d) with a $L$-variate distribution $F_g$.   Assume that the distribution $F_g$  admits finite mean vector $\mu_g$  and positive definite dispersion matrix $\Sigma_g$ (not necessarily multi-normal). 
Following the approach of \cite{Sen2006} and \cite{Pinheiro2009}, we define the functional distance $\theta(F_g,F_h)$ as
\begin{equation}\label{eq_phi_kernel}
 \theta(F_g,F_h)=\int\int\phi(x_1,x_2)dF_g(x_1)dF_h(x_2),
\end{equation}
for an order $2$ symmetric kernel $\phi(\cdot,\cdot)$. If we assume that $\theta(\cdot,\cdot)$ is a convex linear function of its marginal components, then we have
\begin{equation}\label{eq_uneq}
\theta(F_1, F_2) \geq \frac12 \,\{\theta(F_1, F_1) +\theta(F_2, F_2)\},
\end{equation}
for all distributions $F_1$ and $F_2$, where equality sign holds whenever $\mu_1= \mu_2$.

Note that the functional $\theta(\cdot,\cdot)$ can be used to define both distance within and between groups. It follows from U-statistics theory that an unbiased estimator of this functional for within group distance $\theta(F_g,F_g)$ is the $g$-th generalized U-statistics \citep{Hoeffding1948}, with kernel $\phi( \cdot, \cdot)$, defined as

\begin{equation}\label{UnWithin}
 U_{n_g}^{(g)}=\dbinom{n_g}{2}^{-1}\sum_{1\leq i< j\leq n_g}\phi({\bf
X}_{gi},{\bf X}_{gj}).
\end{equation}
Analogously, the unbiased estimator for the between group functional distance $\theta(F_1,F_2)$ is  defined by  
\begin{equation}\label{generalUstat}
 U_{n_1,n_2}^{(1,2)}=\frac{1}{n_1n_2}\sum_{i=1}^{n_1}\sum_{j=1}^{n_2}\phi({\bf
X}_{1i},{\bf X}_{2j}).
\end{equation}
Note that the equivalent combined sample U-statistic can be decomposed as 
\begin{eqnarray}\label{StatisticsUn}
 U_{n} &=& \dbinom{n}{2}^{-1}\sum_{1\leq i< j\leq n}\phi({\bf
X}_i,{\bf X}_j)\nonumber \\
&=&
  \sum_{g=1}^{2} \frac{n_g}{n}U_{n_g}^{(g)}+
\frac{n_1n_2}{n(n-1)}(2U_{n_1 n_2}^{(1,2)}-U_{n_1}^{(1)}-U_{n_2}^{(2)}) = W_n+B_n.
\end{eqnarray}

Decomposition \eqref{StatisticsUn} leads to an essential statistics $B_n$, which provides the theoretical focal point of this paper,   
\begin{eqnarray}\label{StatisticsBn}
 B_{n}&=&
\frac{n_1n_2}{n(n-1)}(2U_{n_1n_2}^{(1,2)}-U_{n_1}^{(1)}-U_{n_2}^{(2)}).
\end{eqnarray}
\noindent Here $U_{n_1}^{(1)}$ and $U_{n_2}^{(2)}$ are U-statistics associated to within group distances,
as defined in \eqref{UnWithin}, and $U_{n_1n_2}^{(1,2)}$ is the U-statistic associated to between group distances as defined in \eqref{generalUstat}. Note that the definitions of $U_{n_1}^{(1)}$ and $U_{n_2}^{(2)}$ require a minimum of 2 elements in the groups. This imposes minimum group sizes $n_1,n_2\geq 2$ for proper definition of $B_n$.

The U-test for group separation considers the issue of verifying whether $G_1$ and $G_2$ in fact constitute separate groups, or if they come from the same distribution. The null hypothesis states that $F_1=F_2$, while the alternative states that they differ. Under the null hypothesis, we have $\E(B_n)=0$ and under the alternative, $\E(B_n) \geq 0$.
The null hypothesis is rejected for large values of standardized $B_n$, where the variance of $B_n$, under the null hypothesis, is obtained by a resampling procedure. 
 The asymptotic properties of $B_n$ are addressed in \citet{Pinheiro2009}. Asymptotic normality is obtained by showing that $B_n$ is in the class of degenerate U-statistics and the convergence rates is $L$ and/or $\sqrt{n}$.

The U-test approach can be extended to assess overall group homogeneity, by verifying whether there exists some significant partition $\{G_1,G_2\}$  of the data. \citet{Valk2012} propose a combinatorial procedure in which a U-test is applied for each possible partition of all group elements into two subgroups. If there is at least one arrangement for which the null hypothesis of group homogeneity is rejected, then the group is considered non-homogeneous. This procedure can only be applied if the group has at least $4$ elements, since we can only consider arrangements where each subgroup has at least two elements. The number of possible assignments of all $n$ elements in 2 subgroups  is $2^{n-1}-n-1$, which becomes an important computational issue, especially for large sample size $n$. To address this issue, \cite{Cybis2016} propose an optimization procedure to assess group homogeneity, by finding the group configuration $G_1$ and $G_2$ that minimizes the objective function
\begin{equation}\label{BnPadrTex}
 f(G_1,G_2)=\frac{-B_n}{\sqrt{\var(B_n)}}.
\end{equation}

 By minimizing this objective function we find  the maximum standardized $B_n$, and thus we must apply  only one test. If this partition is found significant, then the whole group is considered heterogeneous. However, if we do not reject $H_0$ for this partition, then all other partitions will also be non-significant, and the whole group will be considered homogeneous.

Note that while this optimization procedure allows for the application of only one statistical test, the underlying homogeneity criteria is combinatorial, and thus elicits multiple testing. Consequently, we cannot simply apply the U-test to the maximal partition from \eqref{BnPadrTex}. As an alternative approach, we make the simplifying assumption that the $B_n$'s are independent for different group configurations. It is straightforward to show that the asymptotic distribution function of the maximum standardized $B_n$ is given by
\begin{equation}\label{eqtestmaxBn}
 F_{\mbox{max}}(x)= \P\left(\mbox{max}\left(\frac{B_n}{\sqrt{\var(B_n)}}\right)<x\right)=\Phi(x)^{n^*},
\end{equation}
where $\Phi(\cdot)^{n^{*}}$ is the standard normal distribution function at the power $n^*= 2^{n-1}-n -1$. 
If $F_{\mbox{max}}(x)>1-\alpha$, then we reject the null hypothesis of overall group homogeneity with  significance level $\alpha$. The whole group is considered non-homogeneous if and only if we reject $H_0$ in the max-test \eqref{eqtestmaxBn} for this configuration.

Explicit derivations in \citep{Cybis2016} highlight that, under $H_0$, the variance of $B_n$ is different for each subgroup size $n_1 \in \{2, 3,\dots, n-2\} $ , and can be expressed as 

\begin{equation}\label{eq_Var_Bn_text}
 \var(B_n)=  \frac{n_1n_2}{n^2(n - 1)^2}\left[\frac{2n^2-6n+4}{(n_1-1)(n_2-1)}\right]\sigma^4=
 C(n, n_1)\sigma^4,
\end{equation}
where $\sigma^4$ depends only on the covariance structure of
the i.i.d. vectors  $X_1,\cdots,X_n$. Taking advantage of this relation, the optimization algorithm estimates the variance of $B_n$ for one subgroup size through a Monte Carlo permutation procedure, and applies

\begin{equation}\label{varBnijTex}
\widehat{\var_j(B_n)}= \frac{C(n, j)}{C(n, i)}\widehat{\var_i(B_n)},
\end{equation}

\noindent to obtain the variance for other values of $n_1$. Here $\widehat{\var_i(B_n)}$ is an estimate of $B_n$'s variance for $n_1=i$.

\subsection{Extension of Bn}

We wish to explore the homogeneity  method presented in \cite{Cybis2016} to build a hierarchical clustering algorithm.
However, this method is constrained to cases when both subgroups have sizes $n_i\geq 2$, and a hierarchical clustering method should not have this restriction. The group size restriction is a consequence of the definition of the U-statistic $B_n$ from a subgroup decomposability argument, resulting in weighted sums of distances between and within clusters. 

In order to build a clustering algorithm that considers groups of size 1 in the framework of \citet{Cybis2016}, we propose an extension of $B_n$. Define

\begin{equation}\label{Bn_1}
B_n=\left\{
\begin{array}{ll}
\frac{n-1}{n(n-1)}(U_{1,n-1}^{(1,2)}-U_{n-1}^{(2)}) & \mbox{if } n_1 = 1,\\
&\\
\frac{n_1n_2}{n(n-1)}(2U_{n_1n_2}^{(1,2)}-U_{n_1}^{(1)}-U_{n_2}^{(2)}) & \mbox{if }2\leq n_1 \leq n-2,\\
&\\
\frac{n-1}{n(n-1)}(U_{1,n-1}^{(1,2)}-U_{n-1}^{(1)}) & \mbox{if } n_1 = n-1.
\end{array}
\right.
\end{equation}

\noindent We will show that this is a natural extension of $B_n$ when allowing for clusters of size 1.
This extension coincides with that of expression (\ref{StatisticsBn}) for group of sizes $2\leq n_1\leq n-2$, and thus all properties mentioned above are still valid for the new definition. We ascertain the validity of these properties or analogous alternatives in the case of $n_1=1$.

Note that when $n_1=1$ we have 
\begin{eqnarray}\label{eq3_4Pinheiro2009}
U_n &=& \binom{n}{2}^{-1}\sum_{1\leq i <j\leq n}\phi(X_i,X_j)\nonumber\\
&=& \frac{n-1}{n}U_{n-1}^{(1)}+\frac{1}{n}\left(U_{1,n-1}^{(1,2)}-U_{n-1}^{(1)} \right)\nonumber\\
&=& W_n+B_n,
\end{eqnarray}
\noindent where $U_{1,n-1}^{(1,2)}$ and $U_{n-1}^{(1)}$ are as defined in (\ref{UnWithin}) and (\ref{generalUstat}). Thus, $B_n$ still arises from the decomposition of the combined sample U-statistic into $B_n$ and a term $W_n$ which is the mean of within group distances.

This extended definition can be used to build a U-test in the same context of \citet{Cybis2016}. Let $G_1=\{X_1\}$ and $G_2=\{X_2,\dots,X_n\}$, that is $X_1$ is the only sample in group 1 and all other samples are in group 2.
Under the null hypothesis of overall group homogeneity, for $n_1=1$, we still have $\E[B_n]=0$.  Under the alternative, it is natural to require 
\begin{equation}\label{Assump1}
\E(\phi(X_1,X_j))>\E(\phi(X_i,X_j)), \quad \mbox{ for } X_i,\, X_j\in G_2,
\end{equation}
and thus  $\E[B_n]>0$. Note that this assumption is compatible with the case of $n_1\geq 2$ since when \eqref{Assump1} is valid then equation (\ref{eq_uneq}) is always satisfied.

For $2\leq n_1\leq n-2$, the statistic $B_n$ is a degenerate U-statistic and asymptotic normality is established in \cite{Pinheiro2009}. When $n_1=1$ the Hoeffding decomposition shows that $B_n$ is non-degenerate. The following Theorems establish the asymptotic distribution of the extended $B_n$ under $H_0$ for increasing dimension $L$ and sample size $n$, requiring regularity conditions akin to those of \cite{Pinheiro2009}.

\begin{theorem}\label{Theorem1}\rm
Let $X_1, X_2,\dots,X_n$ be a sequence of i.i.d. $L \times 1$ random vectors with a distribution $F$. Let $\phi(\cdot,\cdot)$ be a kernel of degree 2 satisfying $\E[\phi(X_1,X_2)^2]<\infty$ and $\var[\E(\phi(X_1,X_2)|X_1)]=\sigma_1^2>0$. Let $W$ be the distribution of the standardized first therm in the Hoeffding decomposition of the kernel $\phi(\cdot,\cdot)$,  $\psi_1(X_1)/\sqrt{\var(\psi_1(X_1))} \sim W$. Consider the decomposition in \eqref{Bn_1} for the case where $n_1=1$. Then 
\begin{equation}\label{convBn_n}
V_n^{-1/2}B_n\xrightarrow{D} W \quad \mbox{ as } \quad n \rightarrow \infty,
\end{equation}
where $V_n=\var(B_n)$. 
\end{theorem}

Proof: See supplementary material.
\begin{theorem}\label{Theorem2}\rm
Let $X_1, X_2,\dots,X_n$ be a sequence of i.i.d. $L \times 1$ random vectors. Let $\phi(\cdot,\cdot)$ be a kernel of degree 2 such that
\begin{equation}\label{phistar}
\phi(X_i,X_j)=\frac{1}{L}\sum_{\ell=1}^{L}\phi^*(X_{li},X_{lj})
\end{equation}
for some kernel $\phi^*(\cdot,\cdot): \R^2\rightarrow \R$.   Define $\phi_1^*(x_{li})=\E[\phi^*(X_{li},X_{lj})|X_{li}=x_{\ell i}]$ and suppose $\var(\phi_1^*(X_{li}))>0$ and $\var(\phi^*(X_{\ell i},X_{\ell j}))<\infty $. Let $B_n$ be defined by \eqref{Bn_1} for the case where $n_1=1$, and assume that all conditions in Theorem \ref{Theorem1} hold.   Suppose also that
\begin{equation}\label{covcond1}
\sum_{1\leq\ell<m\leq n}^{L}\E[\phi^*(X_{li},X_{lj})\phi^*(X_{mi},X_{mj})]=O(L) \quad \mbox{ as } \quad L\rightarrow \infty
\end{equation}
and 
\begin{equation}\label{covcond2}
\sum_{1\leq\ell<m\leq n}^{L}\E[\phi_1^*(X_{li})\phi_1^*(X_{mj})]=O(L) \quad \mbox{ as } \quad L\rightarrow \infty.
\end{equation}
Then 
\begin{equation}\label{convBn_nL}
V_n^{-1/2}B_n\xrightarrow{D} N(0,1) \quad \mbox{ as } \quad  L\rightarrow \infty.
\end{equation}
\end{theorem}

Proof: See supplementary material.

Definition (\ref{Bn_1}) can also be employed in the homogeneity test of \citet{Cybis2016} allowing for groups of size 1. 
Theorem 2 is central to our development in the HDLSS scenario, where $L$ is large, since the homogeneity test relies heavily on asymptotic normality of $B_n$.  Efficient implementation of the homogeneity test under the Euclidean distance also requires an expression for the variance of $B_n$.   
Exploring the Hoeffding decomposition to obtain this  variance under $H_0$, when $n_1=1$, yields
\begin{equation}\label{Var_Bn_1}
\var(B_n)=\left\{
\begin{array}{ll}
\frac{n_1n_2}{n^2(n - 1)^2}\left[\frac{2n^2-6n+4}{(n_1-1)(n_2-1)}\right]\sigma^4 & \mbox{if }2\leq n_1 \leq n-2,\\
&\\
\frac{(n^2-n+1)}{n^2(n-1)^2}\mu^4+\frac{n^2+2n-4}{n^2(n-1)^2(n-2)} \sigma^4& \mbox{if } n_1 = 1 \mbox{ or } n-1,
\end{array}
\right.
\end{equation}
where $\sigma^4$ and $\mu_4$ are central moments of the i.i.d. vectors $X_1,\dots,X_n$, depending only on the data, and are defined in the supplementary material. Since this variance cannot be written as in expression (\ref{eq_Var_Bn_text}), it cannot be obtained from the variance of other groupings by exploring the relation in (\ref{varBnijTex}). Thus, when extending the homogeneity test of \citet{Cybis2016} to consider groups of size 1, we must perform two different Monte Carlo procedures, one for groups of size $n_1=1$ and one for groups of size $n_1=n/2$ which is then used to obtain the variance for all other group sizes. 

The resampling procedure for estimating the variance of $B_n$ under $H_0$ employs the same concept as the permutation test. It resamples the vectors $X_1,\dots,X_n$ into group $1$ of size 1 and group $2$ of size $n-1$, computing $B_n$ for each iteration. Under $H_1$, when the groups have a high degree of separation, the iteration that puts the original (correct) element $X_1$ in group $1$ and all other elements in group 2  produces a much higher value for $B_n$ than all other rearrangements. For moderate values of $n$, this largely inflates the Monte Carlo variance estimate, thus leading to small values for standardized  $B_n$ and small power. To address this issue, we employ a robust estimator based on quantiles \citep{Ma_and_Genton2000} which leads to much smaller variance estimates in these extreme separation situations. Importantly, the probability of type I error is almost not affected (see supplementary Table S2) since, under $H_0$, this estimator underestimates only slightly the variance. Additionally, in supplementary Table S1 we show a comparison between standard and robust estimators.     
This situation also arises under definition (\ref{StatisticsBn}) when $n$ is small (typically $n\leq5$). In these situations, we also employ the robust estimator.

The homogeneity test, as presented in \citep{Cybis2016}, employs the distribution of the maximum of the $2^{n-1}-n-1$ possible U-tests to assess homogeneity. In order to consider extension \eqref{Bn_1}, we must account for the $n$ additional U-tests for groupings with $n_1=1$, thus, in expression \eqref{eqtestmaxBn} we now have $n^{*}= 2^{n-1}-1$.

A brief simulation study evaluating type I error and power of the U-test when $n_1=1$ can be found in section  S2.2 of the supplementary material.

\subsection{Gumbel's approximation}

As discussed above, the procedure to test group homogeneity is based on the distribution of the maximum of $n^*=2^{n-1}-1$ standard normals. Due to the combinatorial nature of our approach, the number of tests increases rapidly, even for moderate sample size. For example, with $n=10$ samples we have $n^{*}=511$ tests, with $n=30$ samples there are  $n^{*}\approx 5\times 10^8$ tests, and for $n=50$ we have impressive $n^{*}\approx 5\times 10^{14}$ tests.
The maximum distribution in \eqref{eqtestmaxBn} adequately accounts for multiple testing for reasonably small values of $n^*$. However, as $n^*$ rapidly increases this approach has some shortcomings and we gain by exploring extreme value   theory.

Extreme value theory states that the maximum of a sample of i.i.d random normals, after convenient normalization, converges to the Gumbel distribution. Specifically, if $Y_1,Y_2,\dots,Y_m$ is an i.i.d standard normal sequence of random variables and $M_m=\max(Y_1,Y_2,\dots,Y_m)$, \cite{Gnedenko1992}  shows that  
\begin{equation}\label{gumbel1}
P\left(a_m^{-1}(M_m-b_m)\leq y\right)=\exp(-\exp(-y)),\quad \forall \quad y \in \R,
\end{equation}
for appropriate values of $a_m$ and $b_m$. Here we take 
\begin{equation}\label{bn_gumbel} a_m= \frac{\log \left(\left(4 \log^2(2)\right)/\left(\log^2\left(\frac{4}{3}\right)\right)\right)}{2\sqrt{2\log (m)}}\quad \mbox{and} \quad 
b_m=\sqrt{2\log (m)}-\frac{\log (\log (m))+\log \left(4 \pi  \log ^2(2)\right)}{2 \sqrt{2\log (m)}}.
\end{equation}

The homogeneity test statistic is the maximum of the standardized $B_n$'s for all possible group arrangements, which are all asymptotically normal, thus we have $m=n^*$. 
However, the Gumbel approximation is only valid for very large values of $n^*$. Thus, for small $n$ we employ the standard max distribution of \eqref{eqtestmaxBn}, and when $n\geq 30$ the Gumbel distribution. The simulation studies in Section \ref{Sec_Sim} assess statistical properties of the Gumbel approximation and present the comparisons used to define this threshold.

\subsection{Finding Significant Clusters (UClust)}
\label{sec_U-clust}

\begin{figure}[h!]
\begin{center}
\includegraphics[scale=0.3]{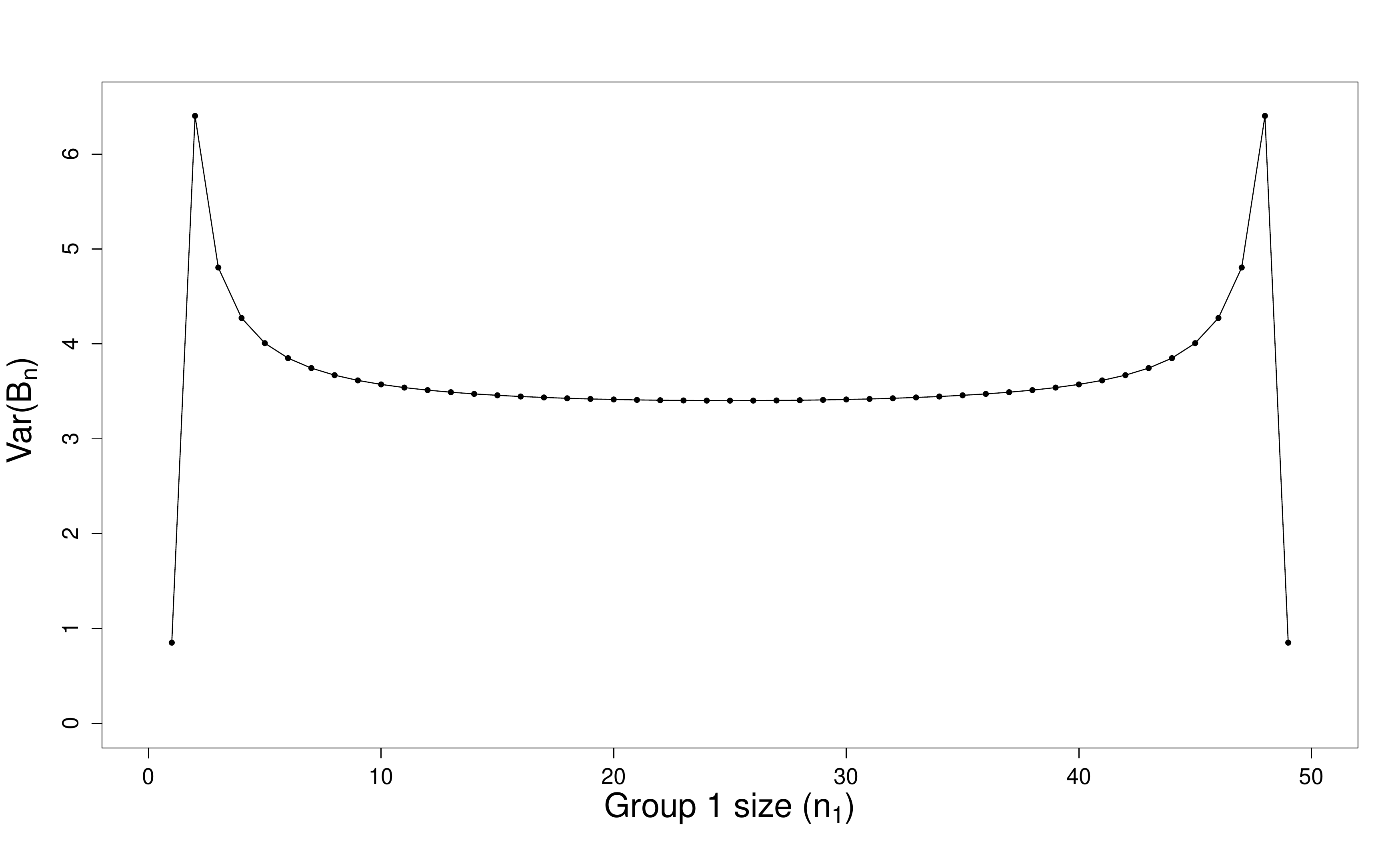} 
\caption{Variance of $B_n$ for different groups sizes $n_1\in\{1,\dots,n-1\}$.}
\label{fig_Smile}
\end{center}
\end{figure}

In order to build the hierarchical clustering procedure, we require a method that finds the statistically significant  partition that better divides a group of samples into two subgroups, if such partition exists. The algorithm employed in the homogeneity test of \citet{Cybis2016} finds a candidate for such partition by maximizing $\frac{B_n}{\sqrt{Var(B_n)}}$. This is appropriate for the homogeneity test, since if the U-test accepts the null hypothesis of homogeneity for this partition, then the null would also be accepted for all other partitions. We note, however, that when more than one significant group separation exists, then the standardized $B_n$ might not be the best criteria to choose between competing partitions. This arises form the fact that the variance of $B_n$ has different magnitudes depending on subgroup size $n_1$. Equation \eqref{Var_Bn_1} dictates a pattern for the relationship between variances, which is shown in figure \ref{fig_Smile}. Consequently, this criteria favours partitions with group sizes of smaller variance, namely  $n_1\approx n/2$ and more markedly $n_1=1$. Table S7 shows how dramatic this effect is when group separation is large. For example, when the data are simulated considering a subgroup of size two, the configuration that maximizes the standardized $B_n$ is always one with a subgroup of size 1.

Considering this effect, we adopt the maximum $B_n$ as the criteria for finding the configuration that better divides the sample into two groups. This approach has an intuitive appeal, since it finds the partition for which between group distances are the largest in relation to within group distances. 
Thus, our significance clustering algorithm will find the partition with maximum $B_n$ among the universe of all significant partitions. If no significant partition exists, then the algorithm will return ``homogeneous". This addresses the issue of group size bias while guaranteeing that the chosen configuration is statistically significant (see Table S7). 

Building an efficient clustering algorithm under this criteria is not straightforward  since we do not wish to exhaustively list all significant partitions. The main issue is that we cannot simply perform a test for the partition that maximizes $B_n$, since there are non-homogeneous groups in which this maximal partition is not significant.  However,  note that the variance of $B_n$ is constant for all partitions with fixed subgroup sizes $n_1$, and that the relationship between the variances for different subgroup sizes is dictated by expression \eqref{Var_Bn_1}. Thus, the partition that maximizes (non standardized) $B_n$ for subgroup size $n_1$ is also the partition that maximizes the standardized $B_n$ for this $n_1$. Consequently, there is no need to search for more than one partition within each subgroup size. Additionally,  if this partition maximizes the standardized $B_n$ in a give universe, then it will also maximize $B_n$ for all subgroup sizes in this universe with variance larger or equal to that of $n_1$.  By exploring this insight, we built the following clustering algorithm based on restricted optimization problems.

\begin{center}
\begin{tabular}{l}
\hline
\noindent{\bf UClust Algorithm}: Finds the data partition that maximizes $B_n$ in the universe of all significant partitions\\ 
\hline
\noindent{\bf Input}: Data X\\
\noindent{\bf Output}: Partition $\{S_1^*,S_2^*\}$\\

1: Apply homogeneity test to X\\
2: {\bf if} Accept $H_0$\\
3: \quad {\bf Return} $S_1^*=\emptyset$ and  $S_2^*=\{1,\dots,n\}$ \\
4: {\bf else}  \\
5:\quad\quad find $S_1^*$ and $S_2^*$ that optimize $B_n$ \\
6:\quad\quad{\bf while}  $\{S_1^*,S_2^*\}$ is not a significant partition\\
7:\quad\quad\quad\quad  find $S_1$ and $S_2$ that optimize $B_n$ for subgroup size $n_1\in \{\#S_1^*,\dots,n-\#S_1^*\}\cup\{1\}$ \\ 
8:\quad\quad\quad\quad  set $S_1^*=S_1$ and $S_2^*=S_2$ \\
9:\quad\quad{\bf Return} $\{S_1^*,S_2^*\}$\\
\hline
\end{tabular}
\end{center}

\subsection{Significance for Hierarchical Clustering (UHClust)}
 
Our hierarchical clustering method is a divisive procedure that consists of sequential application of the UClust method of section \ref{sec_U-clust}. 
At the first level, we start with the whole sample $G_0=(X_1,...,X_n)$ and partition it into two subgroups, $G_1$ and $G_2$, if the sample is not homogeneous. 
Then for each group $G_i$, we once again apply UClust and divide the group into two new subgroups $G_{l}$ and $G_{k}$, whenever $G_i$ is found to be non-homogeneous.   
This procedure is repeated for each new group $G_i$ until all groups are considered homogeneous by UClust or we reach the minimum size $\tau$ stopping criteria. For all the simulations and applications in this paper we set $\tau=3$, thus groups of size 3 or smaller were deemed to small to be further subdivided by UClust.  

The whole hierarchical clustering procedure consists of sequential homogeneity tests, and thus incurs on issues of multiple testing. To control the family-wise error rate (FWER) at level $\alpha \in (0,1)$, we follow the approach of \citet{Kimes2017} by updating the effective type I error rate $\alpha_i$ for each test. Let $n_i$ be the number of elements of group $G_i$, then the UClust test to partition the group is      performed at level

\begin{equation}
\alpha_i=\alpha\frac{n_i-1}{n-1}.
\end{equation}

Thus, at the first level, the test that divides the whole sample $G_0$ is performed at level $\alpha_0=\alpha$, since $n_0=n$. The subsequent tests are performed at decreasing levels $\alpha_i$ since group sizes $n_i$ decrease throughout the hierarchical clustering procedure.

\section{Simulation}\label{Sec_Sim}

\subsection{Gumbel's approximation}
In Section 2.3 we explore extreme value theory to approximate the distribution of the maximum of $B_n$  for large values of $n$.  To evaluate the statistical properties of the homogeneity test considering both the Max distribution of \eqref{eqtestmaxBn} and the Gumbel distribution of \eqref{gumbel1}  we simulate data for $n\in \{10,20,30,40,50,70,100\}$ independent multivariate normally distributed vectors of length $L\in \{500,1000,2000\}$ divided in 2 groups of size $n/2$. The elements in group 1 were simulated with zero mean and identity matrix covariance. The elements of group 2 have the same covariance matrix, and mean vector with all entries equal to $m_2 \in \{0.00, 0.25, 0.5\}$. For each simulated dataset, we applied the max test (M) and the Gumbel corrected  max test (G).
We generate $Re=1000$ replications for each scenario.

Table \ref{Tab_Gumbel} presents the fraction of tests for which the null hypothesis was rejected in each case. 
The lines with $m_2=0.00$ represent estimates of type I error rates, and the remaining lines represent power estimates. 
Note that, while the Max test is generally more powerful, the Gumbel correction allows for better control of type I error, when $n$ increases.  
Simulations for $n=30$ show that around this sample size, the Max test ceases to adequately control the error rate, indicating that for this value of $n$ the Gumbel correction starts being a better option.  
In order to define the precise threshold value of $n$ from which the Gumbel approximation should be used, we performed a series of simulation studies presented in the supplementary material. From Tables S4, S5 and S6  we elect to use the Gumbel correction for $n\geq 30$.

\begin{table}[h!]
\centering
\begin{tabular}{|c|c|cc|cc|cc|cc|cc|cc|cc|}
                
  \hline
      &  n    &\multicolumn{2}{c|}{10}&\multicolumn{2}{c|}{20}&\multicolumn{2}{c|}{30}&\multicolumn{2}{c|}{40}&\multicolumn{2}{c|}{50}&\multicolumn{2}{c|}{70}&\multicolumn{2}{c|}{100}\\
      \hline
L     &$m_2$& M  & G  & M  & G  & M  & G  & M  & G  & M  & G  & M  & G  & M  & G  \\ 
  \hline
      & 0.00 & 0.04 & 0.00 & 0.05 & 0.02 & 0.09 & 0.02 & 0.12 & 0.03 & 0.26 & 0.09 & 0.36 & 0.20 & 0.70 & 0.49 \\ 
  500 & 0.25 & 0.08 & 0.01 & 0.44 & 0.24 & 0.83 & 0.73 & 0.99 & 0.98 & 1.00 & 1.00 & 1.00 & 1.00 & 1.00 & 1.00 \\ 
      & 0.50 & 1.00 & 0.95 & 1.00 & 1.00 & 1.00 & 1.00 & 1.00 & 1.00 & 1.00 & 1.00 & 1.00 & 1.00 & 1.00 & 1.00 \\ 
      \hline
      & 0.00 & 0.03 & 0.03 & 0.03 & 0.01 & 0.05 & 0.00 & 0.08 & 0.03 & 0.06 & 0.02 & 0.11 & 0.06 & 0.34 & 0.23 \\ 
 1000 & 0.25 & 0.15 & 0.02 & 0.87 & 0.79 & 1.00 & 1.00 & 1.00 & 1.00 & 1.00 & 1.00 & 1.00 & 1.00 & 1.00 & 1.00 \\ 
      & 0.50 & 1.00 & 1.00 & 1.00 & 1.00 & 1.00 & 1.00 & 1.00 & 1.00 & 1.00 & 1.00 & 1.00 & 1.00 & 1.00 & 1.00 \\ 
      \hline
      & 0.00 & 0.01 & 0.00 & 0.02 & 0.00 & 0.09 & 0.04 & 0.03 & 0.02 & 0.05 & 0.01 & 0.04 & 0.00 & 0.09 & 0.04 \\ 
 2000 & 0.25 & 0.58 & 0.10 & 1.00 & 0.99 & 1.00 & 1.00 & 1.00 & 1.00 & 1.00 & 1.00 & 1.00 & 1.00 & 1.00 & 1.00 \\ 
      & 0.50 & 1.00 & 1.00 & 1.00 & 1.00 & 1.00 & 1.00 & 1.00 & 1.00 & 1.00 & 1.00 & 1.00 & 1.00 & 1.00 & 1.00 \\ 
   \hline
\end{tabular}
\caption{Proportion of rejection at level $\alpha=5\%$ of the homogeneity test for the Max Distribution (M) and Gumbel correction (G).}
\label{Tab_Gumbel}
\end{table}

The Gumbel correction improves the method's type I error control. This is particularly the case for HDLSS settings, in which $L>>n$. In situations with small $L$ to $n$ ratio, the test struggles at size control.

\subsection{Finding Significant Clusters}

\begin{figure}[h!]
\includegraphics[scale=0.5]{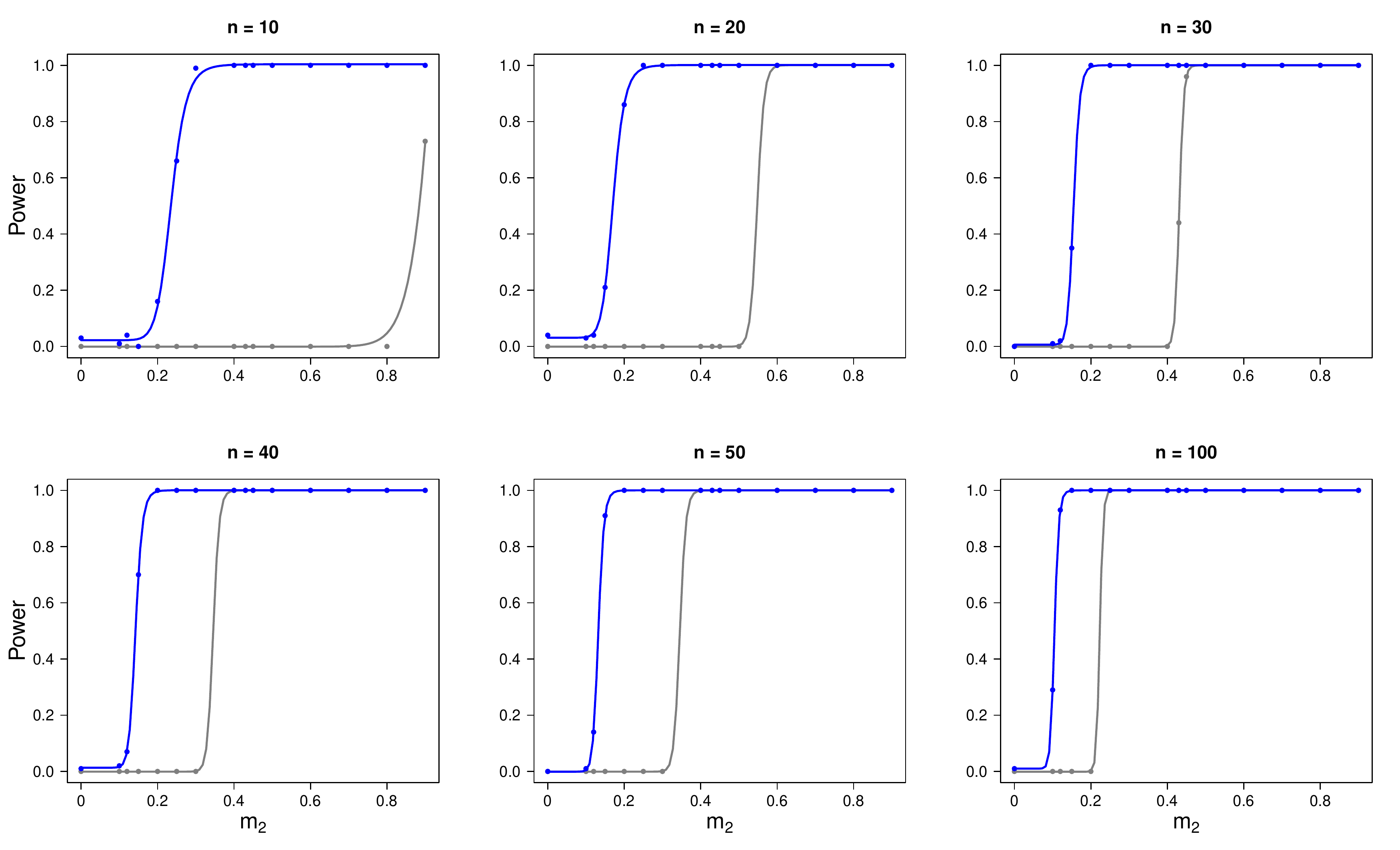} 
\caption{Power curves of Uclust (blue) and SigClust (gray) methods for 100 replications of each scenario with L=2500 and $n_1=n/2$.
}
\label{fig_powercurves1}
\end{figure}

\begin{figure}[h!]
\includegraphics[scale=0.5]{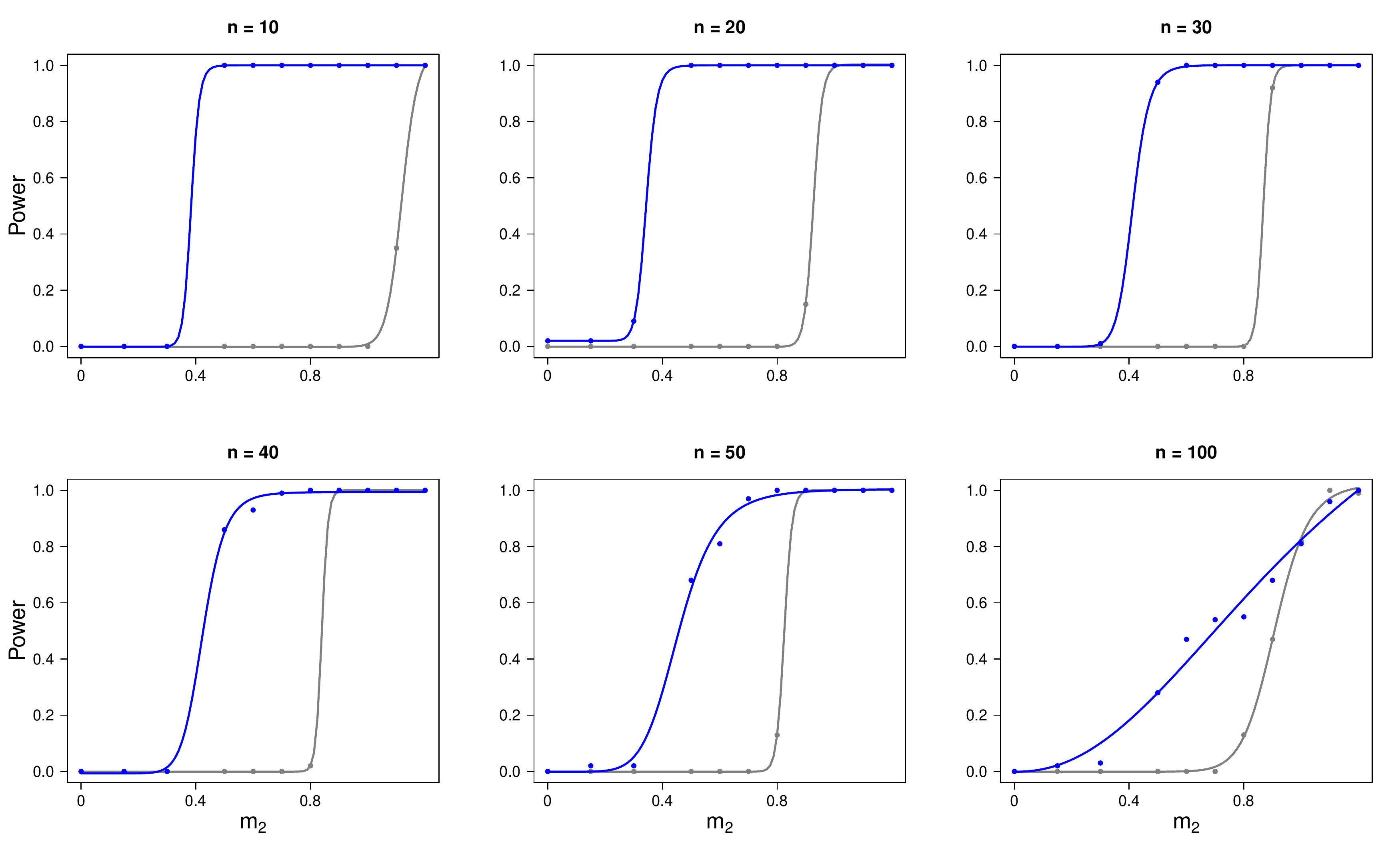} 
\caption{Power curves of Uclust (blue) and SigClust (gray) methods for 100 replications of each scenario with $L=2500$, $n_1=n-2$.
}
\label{fig_powercurves2}
\end{figure}

In order to evaluate our signifcance clustering method, we present simulation studies comparing Uclust to SigClust. The data were simulated under the i.i.d normal model, with $n\in \{10,20,30,40,50,70,100\}$,  $L=2500$, $Re=100$. The elements in group 1 were simulated with zero mean and identity matrix covariance. The elements of group 2 have the same covariance matrix, and varying mean vectors.
Figures \ref{fig_powercurves1} and \ref{fig_powercurves2} present power curves for both methods, respectively for $n_1=n/2$ and $n_1=2$.  More details of these results are presented in Supplementary Tables S8 and S9. Both method have adequate control of type I error for most scenarios and Uclust is generally more powerful than SigClust \citep{Liu2008} across all simulation scenarios. The control of type I error and the superior power can also be observed when non-normal models are considered.
Tables S10 and S11 show simulation results for two non-symmetric models,  Chi-squared  and Gamma. Again we observed a favourable performance in terms of power for Uclust when compared to SigClust.

Both figures show that Uclust has more statistical power than SigClust in all scenarios. The difference is particularly pronounced when samples size $n$ is small or  when group are unequally balanced. When groups have the same size, increases in sample size are associated with increases power, although this effect is less marked for Uclust. Additionally, as expected, when groups are unbalanced, both test have lower power.

\FloatBarrier

\subsection{Hierarchical Clustering}

We present here a simulation  study to evaluate the performance of Uhclust in comparison with competing methods pvclust \citep{suzuki2006} and Shclust \citep{Kimes2017}. The data were simulated from multivariate normal vectors of length $L=2500$ with identity covariance in  $K\in \{3, 7\}$ clusters  for varying degrees of cluster separation  $d$ and varying sample sizes $n= n_1 \times K$. 
Two simulation scheme were considered. Table \ref{Tab_Hc_triang} presents results for data simulated according to a scheme in which all cluster means were equidistant, with mean distance between any pair equal to $d$. In table \ref{Tab_Hc_inline}, the data were simulated with all cluster means in a linear arrangement with $d$ distance between one mean and its closest neighbours.  Supplementary Tables S12 and S13 show the analogous results for $K=5$. We run $Re=100$ replications for each setting.

The methods were compared in terms of mean Adjusted Rand Index (ARI) which measures the agreement of clustering results with simulation scenarios, adjusting for randomness. An ARI of one indicates perfect matching. Additionally, we compared the average number of significant clusters found by each method.

All tests were performed at a significance level $\alpha=0.05$. However, while in both Uhclust and Shclust the null hypothesis is stated in terms of group homogeneity, pvclust results cannot be interpreted under the same light. Its p-values are actually probability values for   dendrogram nodes, which are evaluated through the bootastrap. Consequently, some elements might not belong to any significant cluster.

\begin{table}[ht]
\centering
\begin{tabular}{|c|c|ccc|ccc|ccc|ccc|}
\hline
\multicolumn{2}{|c}{}&\multicolumn{6}{|c|}{\small K\normalsize =3}&\multicolumn{6}{c|}{\small K\normalsize =7}\\ 
\cline{3-14}
\addlinespace[1pt]
\multicolumn{2}{|c}{Parameters}&\multicolumn{3}{|c}{mean ARI}&\multicolumn{3}{c|}{mean \small$\widehat{K}$\normalsize}&\multicolumn{3}{c|}{mean ARI}&\multicolumn{3}{c|}{mean \small $\widehat{K}$\normalsize}\\
\hline
$d$&$n_1$& uhc & shc & pvc & uhc & shc & pvc& uhc & shc & pvc & uhc & shc & pvc\\
\hline
     &10   & 0.43& 0.00& 0.06& 2.11& 1.00&  1.48 & 0.95& 0.00& 0.37& 7.08& 1.00& 4.07\\ 
0.2  &20   & 0.89& 0.00& 0.05& 3.07& 1.00&  6.34 & 0.99& 0.72& 0.17& 7.04& 5.38& 2.51\\ 
     &30   & 0.95& 0.03& 0.03& 3.03& 1.06& 13.56 & 0.99& 0.95& 0.19& 7.00& 6.95& 2.94\\ 
 \hline                                                                                 
     &10   & 1.00& 0.14& 0.81& 3.05& 1.25&  3.06 & 1.00& 0.42& 0.87& 7.02& 3.33& 6.37\\ 
0.4  &20   & 0.99& 1.00& 0.93& 3.06& 3.00&  2.92 & 1.00& 1.00& 0.93& 7.05& 6.97& 6.58\\ 
     &30   & 1.00& 1.00& 0.97& 3.01& 3.00&  2.92 & 1.00& 1.00& 0.94& 7.00& 7.00& 6.60\\ 
 \hline                                                                                 
     &10   & 1.00& 0.12& 0.77& 3.00& 1.21&  2.95 & 1.00& 0.47& 0.95& 7.09& 3.77& 6.61\\ 
0.6  &20   & 0.99& 1.00& 1.00& 3.07& 3.00&  3.00 & 1.00& 1.00& 0.98& 7.08& 7.00& 6.88\\ 
     &30   & 1.00& 1.00& 1.00& 3.01& 3.00&  3.00 & 1.00& 1.00& 0.97& 7.00& 7.00& 6.84\\ 
   \hline
\end{tabular}
\caption{ Comparison of mean ARI and mean number of significant clusters $\hat{K}$ for Uhclust (uhc), Shculst (shc) and pvclust (pvc), for equidistantly simulated means.
}
\label{Tab_Hc_triang}
\end{table}


\begin{table}[ht]
\centering
\begin{tabular}{|c|c|ccc|ccc|ccc|ccc|}
\hline
\multicolumn{2}{|c}{}&\multicolumn{6}{|c|}{\small K\normalsize =3}&\multicolumn{6}{c|}{\small K\normalsize =7}\\ 
\cline{3-14}
\addlinespace[1pt]
\multicolumn{2}{|c}{Parameters}&\multicolumn{3}{|c|}{mean ARI}&\multicolumn{3}{c|}{mean \small$\widehat{K}$\normalsize}&\multicolumn{3}{c|}{mean ARI}&\multicolumn{3}{c|}{mean \small$\widehat{K}$\normalsize}\\
\hline
$d$&$n_1$& uhc & shc & pvc & uhc & shc & pvc& uhc & shc & pvc & uhc & shc & pvc\\
\hline
     &10   & 0.50 & 0.00 & 0.13 & 2.11 & 1.00 &  1.88 &  0.59& 0.01& 0.36& 4.43& 1.03& 3.44\\  
0.2  &20   & 0.55 & 0.03 & 0.09 & 2.41 & 1.06 &  7.00 &  0.70& 0.13& 0.26& 5.73& 1.55& 1.79\\  
     &30   & 0.71 & 0.17 & 0.04 & 2.88 & 1.38 & 13.72 &  0.83& 0.17& 0.29& 6.92& 1.74& 1.84\\  
 \hline                                                                                        
     &10   & 0.99 & 0.01 & 0.41 & 3.05 & 1.01 &  2.34 &  1.00& 0.10& 0.41& 7.03& 1.40& 2.98\\  
0.4  &20   & 0.99 & 0.06 & 0.50 & 3.03 & 1.12 &  2.62 &  1.00& 0.35& 0.40& 7.03& 2.92& 2.95\\  
     &30   & 1.00 & 0.42 & 0.61 & 3.01 & 1.84 &  3.37 &  1.00& 0.48& 0.41& 7.00& 4.00& 3.00\\  
 \hline                                                                                        
     &10   & 1.00 & 0.00 & 0.54 & 3.05 & 1.00 &  2.20 &  1.00& 0.20& 0.39& 7.03& 1.79& 2.99\\  
0.6  &20   & 0.99 & 0.02 & 0.83 & 3.07 & 1.04 &  2.59 &  1.00& 0.42& 0.41& 7.07& 3.61& 3.00\\  
     &30   & 1.00 & 0.55 & 0.92 & 3.01 & 2.10 &  2.84 &  1.00& 0.48& 0.41& 7.02& 3.98& 3.00\\  
   \hline
\end{tabular}
\caption{Comparison of mean ARI and mean number of significant clusters $\hat{K}$ for Uhclust (uhc), Shculst (shc) and pvclust (pvc), for inline simulated means.
}
\label{Tab_Hc_inline}
\end{table}

The simulations show that Uhclust is generally superior to the competing methods, both in terms of identifying the correct number of clusters ($\hat{K}$) and in terms of actual element assignment  (ARI). Additionally, as with the other methods, we note that Uhclust is consistent, improving in both metrics with increases of group separation $d$ and group size $n_1$.

\FloatBarrier
\section{Applications}

\subsection{Genetic Diversity of Human Populations}

We consider a dataset from the human genetic diversity project (HGDP) consisting of  377 autosomal microsatellite markers in 1056 individuals from 52 ancestral populations (\citealp{rosenberg2002}). These data have previously been considered in hierarchical settings to assess the genetic relationships between the human populations (\citealp*{rosenberg2002,chen_et_al_2015}). 

\begin{figure}
\includegraphics[scale=0.5]{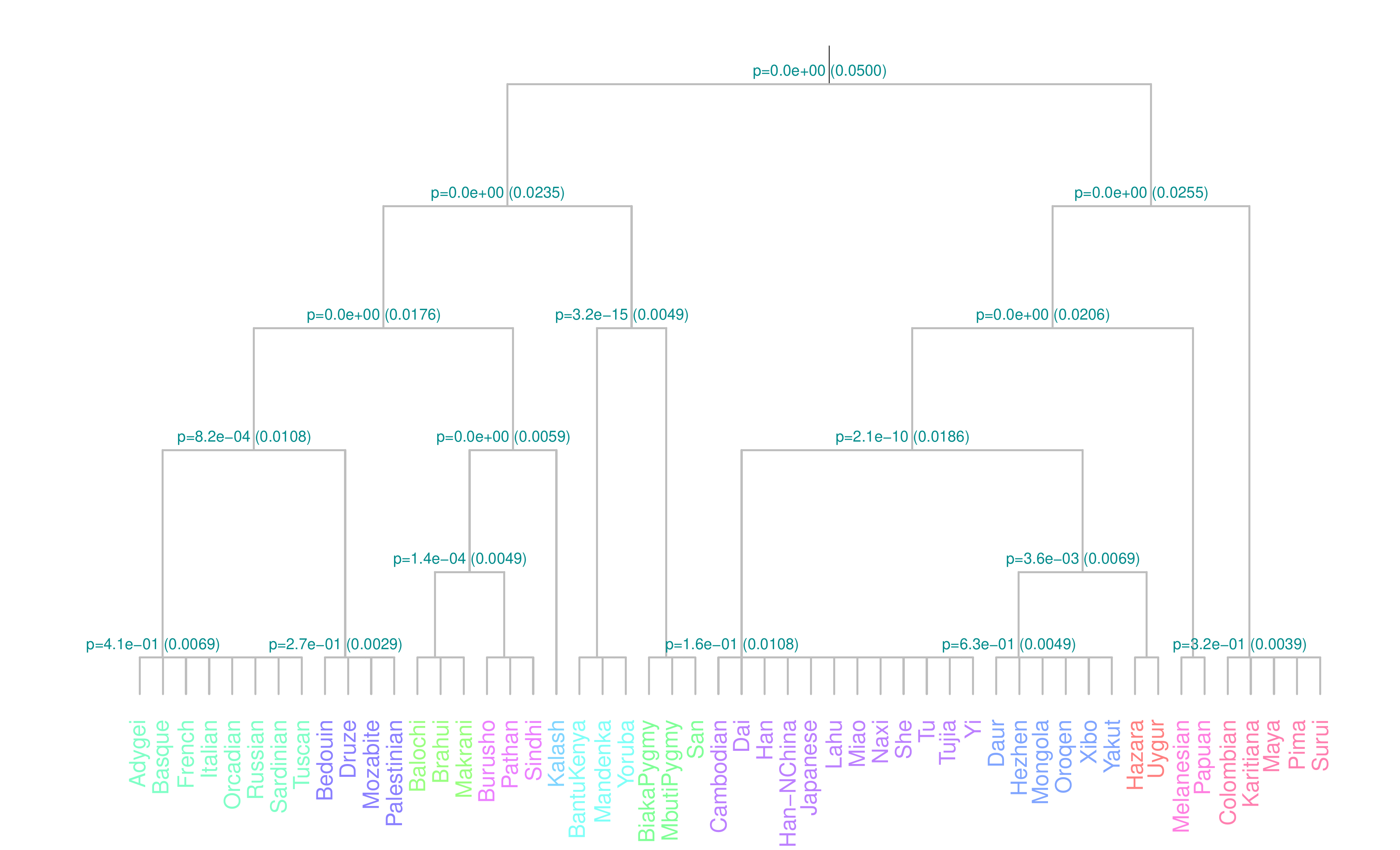} 
\caption{Annotated dendrogram of significance analysis for hierarchical clustering for 52 human populations. P-values and corrected significance levels $\alpha^*$ are shown for each test performed at the corresponding node.}
\label{fig_HGDP}
\end{figure}

We apply our hierarchical clustering algorithm to identify statistically significant population clusters. Figure \ref{fig_HGDP} presents a dendrogram of these results annotated with p-values and corresponding corrected significance levels $\alpha^*$ for each test performed. We found a total of 12 groups, but only 5 of them were homogeneous clusters with 4 or more populations. The first level division partitions the populations into a group consisting mainly of African and Indo-European populations, and a second group dominated by populations from east Asia, Oceania and the Americas. At the second level division, the first group is partitioned into an African group, and the remaining populations, an the second group into an American and the remaining populations. 

It is noteworthy that the American, European and Middle Eastern groups presented homogeneous population clusters. Additionally, we found two larger homogeneous clusters of Central/East Asian populations. However, no homogeneous clusters were found for both African and the Indian/Pakistani populations, indicating greater genetic heterogeneity in these locations.    
A SigClust analysis of the same data found only 2 significant clusters (Supplementary Figure S1).

\subsection{Breast Cancer Gene Expression}

Finally, we consider a dataset of microarray gene expression for 1645 well-chosen intrinsic genes in 337 Breast Cancer samples, originally obtained from the University of North Carolina (UNC) Microarray Database (https://genome.unc. edu/pubsup/clow/) and compiled in \cite{Prat2010}. 
These data have previously been divided into five molecular
subtypes: Luminal A , Luminal B, Basal-like,
Normal breast-like, and HER2-enriched \citep{Parker2009}. 

\begin{figure}
\includegraphics[scale=0.47]{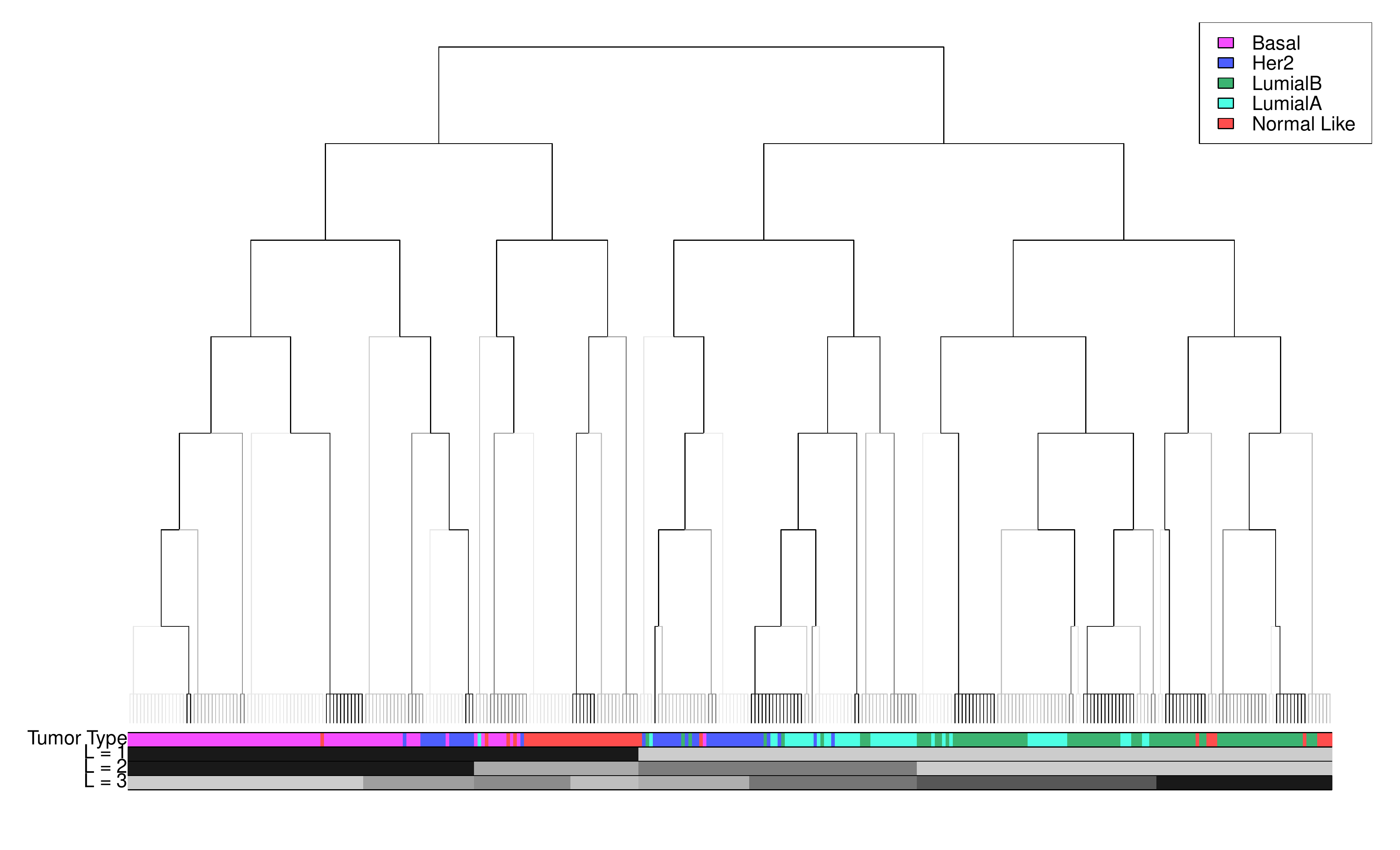} 
\caption{Dendrogram of significance analysis for hierarchical clustering for the BRCA dataset. The coloured bar corresponds to original annotations and the gray bars correspond to the first three levels of significant hierarchical clustering.}
\label{fig_BRCA}
\end{figure}

We apply our hierarchical clustering algorithm to assess weather we can detect these clusters from the molecular data and weather there are genetically homogeneous groups in the  dataset. Figure \ref{fig_BRCA} presents the resulting dendrogram, which showcases the heterogeneous nature of the data. The horizontal bars at the bottom present group separations for the first three levels of hierarchical clustering. The first level partitions the samples into a group mainly consisting of Basal and Normal-like tumours and another formed mainly by the Lumial type. Her2-enriched cells were divided between both groups.      
The second level gives us three groups predominately represented respectively by Basal, Normal-like, Lumial B, and a fourth group mainly consisting of Lumial A and some Her2. The third level divisions still caries some resemblance to these molecular subtypes, however we detect additional levels of genetic heterogeneity that cannot be linked to these subtypes. Final homogeneous cluster sizes range from k=4 to k=23.

These data have been previous analysed through SigClust in \cite{Kimes2017}, in which they find only three significant clusters. Regarding the molecular subtypes, their ARI for these data is 0.42, while ours is 0.15. The Uhclust analysis produces many more clusters than seems relevant considering the predetermined labels, however it showcases additional heterogeneity in the data.

\end{document}